\documentclass[11pt]{article}
\usepackage{moriond,epsfig}

\def\Journal#1#2#3#4{{#1} {\bf #2}, #3 (#4)}

\def\PRL{\em Phys. Rev. Lett.}
\def\PRD{{\em Phys. Rev.} D}

\def\EPJ{{\em Eur. Phys. J.} C}
\def\NAT{\em Nature}
\def\JHEP{\em Journal of High Energy Physics}

\def\be{\begin{equation}}
\def\ee{\end{equation}}
\def\bea{\begin{eqnarray}}
\def\eea{\end{eqnarray}}

\def\calB    {\ensuremath{{\cal B}}\xspace}
\def\CP                {\ensuremath{C\!P}\xspace}
\def\pep2{PEP-II}
\def\VV     {\ensuremath{VV}\xspace} 
\def\VT     {\ensuremath{VT}\xspace} 
\def\VS     {\ensuremath{VS}\xspace} 
\def\Bmeson  {\B\ meson}
\def\Bmesons {\B\ mesons}
\def\fl   {\ensuremath{ f_L}}
\def\ACP  {\ensuremath{A_{\CP}}}

\input{babarsym}

\begin{document}


\vspace*{4cm}

\title{CHARMLESS HADRONIC $B$ DECAYS AT BELLE and \babar}
\author{F.F. WILSON}

\address{Rutherford Appleton Laboratory, Didcot, Chilton, Oxford, OX11
  0QX, UK}

\maketitle\abstracts{
I report on recent measurements from the Belle and \babar\
collaborations on the decay of the \Bmeson\ to hadronic final states
without a charm quark.}

\section{Introduction.}

The study of the branching fractions and angular distributions of
\Bmeson\ decays to hadronic final states without a charm quark probes
the dynamics of both the weak and strong interactions, and plays an
important role in understanding \CP\ Violation (CPV) in the quark sector.
\CP\ Violation at the $B$ factories is described graphically by a
triangle with sides formed from the CKM matrix elements
$V_{qd}V^{\ast}_{qb}$ ($q=u,c,t$) and internal angles
$\alpha,\beta,\gamma$ (or $\phi_2,\phi_1,\phi_3$). Discrepancies in
the measured values of the sides and angles could be an indication of
New Physics beyond the Standard Model (SM) due to enhanced branching
fractions or modified \CP\ asymmetries. The experimental measurements of
branching fractions, \CP\ asymmetries, polarization and phases (both
weak and strong) can be compared to theoretical models based on, for
example, QCD factorization, SU(3) symmetry and Lattice QCD.

The results presented below assume charge-conjugate states and all
branching fraction upper limits (UL) are at the 90\% confidence level
(C.L.). The time-integrated \CP\ asymmetry is defined as $\ACP =
(N_b - N_{\overline{b}}) / (N_b + N_{\overline{b}})$ where $N_b$
($N_{\overline{b}}$) is the number of \Bmesons\ containing a
$b$($\overline{b}$) quark. The latest results are based on a total
dataset of $467\times 10^{6}$ \BB\ pairs for \babar\ and $657\times
10^{6}$ \BB\ pairs for Belle, unless indicated.

\section{Decays involving two-body final states.}
\label{subsec:twobody}

The last few years have seen considerable advancement in the
prediction of the branching fractions and polarizations of \Bmeson\
decays to Vector-Vector (\VV), Vector-Scalar (\VS) and Vector-Tensor (\VT)
final states. In general, there has been good agreement between theory
and experiment on branching fractions (with some notable exceptions) but the polarization
measurements have presented a challenge. The \VV\ states are
expected to be almost fully longitudinally polarized ($\fl \sim 1$)
and this should remain true in the presence of penguin loop
decays. However, penguin-dominated decays seem to have a smaller \fl\
 (e.g. $\fl \sim 0.5$ for $B\to\phi K^{\ast}$)~\cite{bib:p4a}.


Belle has recently measured the decay $\Bm \to K^{\ast0}K^-$ which is
dominated by $b\to d s\bar{s}$ gluonic penguin diagrams. They measure
a yield of $47.7\pm11.1$ events, corresponding to a branching fraction
$\calB (\Bm \to K^{\ast0}K^-) = (0.68\pm0.16\pm0.10) \times 10^{-6}$
with a 4.4$\sigma$ significance~\cite{bib:p4b}. The event yield for
$\Bm \to K_2^{\ast0}(1430)\Km$ is measured to be $23.4\pm12.1$ with an
upper limit on the branching fraction of $\calB (\Bm \to
K_2^{\ast0}(1430)K^-) < 1.1\times 10^{-6}$. A similar analysis has
been done for \Bz\ decays to the \VV\ final states $\rho^0 K^{\ast0}$
and $f_0 K^{\ast0}$~\cite{bib:p7}. Unlike an earlier \babar\
analysis~\cite{bib:p7b}, Belle sees no evidence for $\rho^0 K^{\ast0}$
and $f_0 K^{\ast0}$ (and, consequently, do not measure \fl) but
observes $\Bz\to \rho^0\Kp\pim$ and sees first evidence for $\Bz\to
f_0\Kp\pim$ and $\Bz\to \pip\pim K^{\ast0}$, with branching fractions
(significance) of $(2.8 \pm 0.5 \pm 0.5)\times 10^{-6}$ ($5.0\sigma$),
$(1.4 \pm 0.4^{+0.3}_{-0.4})\times 10^{-6}$ ($3.5\sigma$), and
$(4.5^{+1.1+0.9}_{-1.0-1.6})\times 10^{-6}$ ($4.5\sigma$),
respectively. \babar\ has measured \Bmeson\ decay to an $\omega$
accompanied by a $K^{\ast}$, $\rho$ or $f_0$. Five measurements have
a significance above 5$\sigma$, with another two above 3$\sigma$. This
has allowed \babar\ to measure both \fl\ and \ACP. The \VV\ branching
fractions agree with theory predictions and the asymmetries are
consistent with zero, as expected, while $\fl \sim 0.5$ except for
$\omega\rho^+ \sim 0.9$.  The results~\cite{bib:p8} are summarized in
Table~\ref{tab:vv}.

\begin{table}[!htb]
\caption{Branching fraction central value (\calB) and upper limit (UL)
  in units of $10^{-6}$, significance S in standard deviations, longitudinal polarization
  (\fl) and \CP\ asymmetry A$_{\CP}$ for the Vector-Vector (\VV), Vector-Scalar (\VS) and
  Vector-Tensor (\VT) decays of $B\to\omega K^{\ast}$, $\omega
  f_0$ and $\omega\rho$. 
\label{tab:vv}}
\vspace{0.4cm}
\begin{center}
\begin{tabular}{|c|l|c|c|c|c|c|}
\hline
Mode & Decay & S($\sigma$) & \calB\ & UL & \fl\ & A$_{\CP}$ \\
\hline
\VV & $\omega K^{\ast0}$ & $4.1$ & $2.2\pm0.6\pm0.2$ & - 
   & $0.72\pm0.14\pm0.02$ & $+0.45\pm0.25\pm0.02$ \\
\VV & $\omega K^{\ast+}$ & $2.5$ & $2.4\pm1.0\pm0.2$ & 7.4 
   & $0.41\pm0.18\pm0.05$ & $+0.29\pm0.35\pm0.02$ \\
\VS & $\omega (K\pi)^{\ast0}_0$ & $9.8$ & $18.4\pm1.8\pm1.7$ & - 
   & - & $-0.07\pm0.09\pm0.02$ \\
\VS & $\omega (K\pi)^{\ast+}_0$ & $9.2$ & $27.5\pm3.0\pm2.6$ & - 
   & - & $-0.10\pm0.09\pm0.02$ \\
\VT & $\omega K^{\ast}_2(1430)^0$ & $5.0$ & $10.1\pm2.0\pm1.1$ & - 
   & $0.45\pm0.12\pm0.02$ & $-0.37\pm0.17\pm0.02$ \\
\VT & $\omega K^{\ast}_2(1430)^+$ & $6.1$ & $21.5\pm3.6\pm2.4$ & - 
   & $0.56\pm0.10\pm0.04$ & $+0.14\pm0.15\pm0.02$ \\
\VV & $\omega \rho^0$ & $1.9$ & $0.8\pm0.5\pm0.2$ & 1.6
   & - & - \\
\VV & $\omega f_0$ & $4.5$ & $1.0\pm0.3\pm0.1$ & 1.5
   & - & - \\
\VV & $\omega \rho^+$ & $9.8$ & $15.9\pm1.6\pm1.4$ & - 
   & $0.90\pm0.05\pm0.03$ & $-0.20\pm0.09\pm0.02$ \\
\hline
\end{tabular}
\end{center}
\end{table}

\section{Decays involving three-body final states.}
\label{subsec:threebody}

An interesting use of the decay to final states with three particles
is the search by Belle for the exotic state X(1812) in the decay
$\Bp\to \Kp\ X(1812), X(1812)\to \omega\phi$. This is similar to the
observation by Belle of the Y(3940) resonance in
$\Bp\to\Kp\omega\psi$~\cite{bib:p5a}. Belle observe $N_{\Kp\omega\phi}
= 22.1^{+8.3}_{-7.2}$ events leading to a branching fraction for the
Dalitz plot of $\calB(\Bp\to\Kp\omega\phi) = (
1.15^{+0.43+0.14}_{-0.38-0.13})\times 10^{-6}$ (2.8$\sigma$) and an
upper limit $< 1.9\times 10^{-6}$. Assuming the X(1812) masses and
width from BES~\cite{bib:bes}, Belle searches for a near-threshold
enhancement in the $M_{\pip\pim\piz\Kp\Km}$ mass spectrum.  No
significant yield is seen and an upper limit of $3.2\times 10^{-7}$ is
placed on the product branching fraction $\calB(\Bp\to\Kp X(1812),
X(1812)\to \omega\phi)$~\cite{bib:p5b}.

\babar\ has also looked at rare processes in Dalitz plots. Previous
measurements have shown that almost 50\% of the events in
$\Bz\to\Kp\Km\pip$ can be assigned to an ill-defined resonance,
called $f_X(1500)$ by \babar. If this is an even-spin resonance, the rate for
$f_X(1500)\to \KS\KS$ would be expected to be half the rate for 
$f_X(1500)\to \Kp\Km$. 
They see $15\pm15$ events in the whole
Dalitz plot placing an upper limit on the total branching fraction of
$\calB(\Bp\to \KS\KS\pip) < 5.1\times 10^{-7}$. This makes
the even-spin hypothesis unlikely but interpretation is difficult as
the exact quantum numbers of the $f_X(1500)$ are unknown~\cite{bib:p6a}.

Some MSSM models could enhance the branching fractions of
SM-suppressed decays from the SM values of $\sim10^{-16}$ to
$\sim10^{-6}$. \babar\ has searched for $\Bm\to \Kp\pim\pim$ and
$\Bm\to \Km\Km\pip$ and placed upper limits of $9.5\times10^{-7}$ and
$1.6\times10^{-7}$, respectively, on the branching
fractions~\cite{bib:p6b}.

The decay $\Bp\to\pip\pip\pim$ can in principle be used to extract the
CKM angle $\gamma$ by measuring the interference between $\pip\pim$
resonances and the $\chi_{c0}$ resonance which has no \CP\ violating
phase.  It can also be helpful in understanding broad resonances and
nonresonant backgrounds that are present in $\Bz\to\pip\pim\piz$ and
so improve our measurement of the CKM angle $\alpha$.  \babar's
results~\cite{bib:p9} for $\Bp\to\pip\pip\pim$ are summarized in
Table~\ref{tab:pipipi}.  No significant direct \CP\ asymmetry is
measured and, although some resonances are significant, no evidence is
found for $\chi_{c0}$ and $\chi_{c2}$, leading to branching fraction
upper limits for $\Bp\to\chi_{c0}\pip<1.5\times 10^{-5}$ and
$\Bp\to\chi_{c2}\pip<2.0\times 10^{-5}$, making the measurement of
$\gamma$ in this mode unlikely at Belle or \babar.

\begin{table}[!htb]
\caption{Branching fraction (\calB), \CP\
  asymmetry A$_{CP}$, and Fit Fraction 
for the decay $\Bp\to\pip\pip\pim$ with the resonance decaying to
  $\pip\pim$\label{tab:pipipi}. The errors are
  statistical, systematic and model-dependent, respectively.}
\vspace{0.4cm}
\begin{center}
\begin{tabular}{|l|c|c|c|}
\hline
Decay & Fit Fraction (\%) & \calB ($\times 10^{-6}$)  & A$_{CP}$ (\%) \\
\hline
\noalign{\vskip1pt}
\pip\pip\pim Total & - & $15.2\pm0.6\pm1.2^{+0.4}_{-0.3}$ &
$3.2\pm4.4\pm3.1^{+2.5}_{-2.0}$ \\
\pip\pip\pim nonresonant & $34.9\pm4.2\pm2.9^{+7.5}_{-3.4}$
& $5.3\pm0.7\pm0.6^{+1.1}_{-0.5}$ & $-14\pm14\pm7^{+17}_{-3}$ \\
$\rho^0(770)\pipm;\rho^0\to\pip\pim$ & $53.2\pm3.7\pm2.5^{+1.5}_{-7.4}$
& $8.1\pm0.7\pm1.2^{+0.4}_{-1.1}$ & $18\pm7\pm5^{+2}_{-14}$ \\
$\rho^0(1450)\pipm;\rho^0\to\pip\pim$ & $9.1\pm2.3\pm2.4^{+1.9}_{-4.5}$
& $1.4\pm0.4\pm0.4^{+0.3}_{-0.7}$ & $-6\pm28\pm20^{+12}_{-35}$ \\
$f_2(1270)\pipm;f_2\to\pip\pim$ & $5.9\pm1.6\pm0.4^{+2.0}_{-0.7}$
& $0.9\pm0.2\pm0.1^{+0.3}_{-0.1}$ & $41\pm25\pm13^{+12}_{-8}$ \\
$f_0(1370)\pipm;f_0\to\pip\pim$ & $18.0\pm3.3\pm2.6^{+4.3}_{-3.5}$
& $2.9\pm0.5\pm0.5^{+0.7}_{-0.5}(<4.0)$ & $72\pm15\pm14^{+7}_{-8}$ \\
$f_0(980)\pipm;f_0\to\pip\pim$ & - & $<1.5$ & - \\
$\chi_{c0}\pipm;\chi_{c0}\to\pip\pim$ & - & $<0.1$ & - \\
$\chi_{c2}\pipm;\chi_{c2}\to\pip\pim$ & - & $<0.1$ & - \\
\hline
\end{tabular}
\end{center}
\end{table}

\section{\CP\ Violation and the CKM angle $\alpha(\phi_2)$.}
\label{subsec:cpv}

\begin{figure}[!htb]
\begin{center}
\begin{tabular}{c}
    \epsfig{file=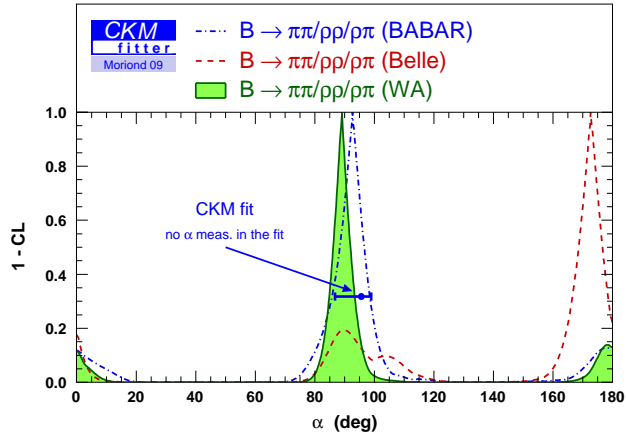,width=0.55\columnwidth}
\end{tabular}
\caption
{Constraints on $\alpha$ ($\phi_2$) from $\B\to\pi\pi$,
  $\B\to\rho\pi$ and 
$\B\to\rho\rho$ \babar\ and Belle measurements compared
  to the prediction from the global CKM fit from CKMfitter. Similar
  results are available from the UTfit group~\protect\cite{bib:ckmfitter}.}
\label{fig:ckmfitter}
\end{center}
\end{figure}

The precision of the measurement of the CKM angle $\alpha(\phi_2)$ continues
to improve. In the absence of penguin loops in the decays, the angle
$\alpha$ can be measured in the time-dependent decay of
$\Bz\to\rho\rho$ and $\Bz\to\pi\pi$. However the penguin contribution,
particularly in $\piz\piz$, is not small and so the measured
$\alpha_{eff}$ differs from the true $\alpha$ by
$\Delta\alpha=\alpha-\alpha_{eff}$. $\Delta\alpha$ can be constrained
by performing an Isospin analysis on the decays $\Bz\to\rho^0\rho^0$,
$\Bpm\to\rho^{\pm}\rho^0$ and
$\Bz\to\rho^+\rho^-$. Table~\ref{tab:rhorho} summarizes the
measurements from \babar~\cite{bib:p10}, where the \CP\ parameters are
quoted for the longitudinally polarized (\CP-even) component of the
\VV\ decays.  When combined, $\Delta\alpha$ is constrained to be
between $-1.8^{o}$ and $6.7^{o}$ (68\% C.L.).  The angle $\alpha$ is
measured to be $(92.4^{+6.0}_{-6.5})^{o}$ and can be compared to the
recent result from Belle~\cite{bib:p10a} of $\alpha =
(91.7\pm14.9)^{o}$. A similar analysis using $B\to\pi\pi$ decays
produces a looser constraint $\mid\Delta\alpha\mid < 43^{o}$, which
results in an exclusion range for $\alpha$ between $23^{o}$ and
$43^{o}$ at the 90\% C.L. The result of combining these measurements
using the CKMfitter programme~\cite{bib:ckmfitter} with earlier
measurements of $B\to\pi\rho$ are shown in Fig.~\ref{fig:ckmfitter}.

\begin{table}[!htb]
\caption{Branching fraction (\calB), longitudinal polarization
  (\fl), direct \CP\ asymmetry (C$_L$),  \CP\ asymmetry in
  the interference between mixing and decay (S$_L$) and \CP\
  asymmetry A$_{CP}$ for the decays  $\Bz\to\rho^+\rho^-$,
  $\Bz\to\rho^0\rho^0$ and $\Bp\to\rho^+\rho^0$ measured by \babar.\label{tab:rhorho}}
\vspace{0.4cm}
\begin{center}
\begin{tabular}{|l|c|c|c|}
\hline
& $\Bz\to\rho^+\rho^-$ & $\Bz\to\rho^0\rho^0$ & $\Bp\to\rho^+\rho^0$ \\
\hline
\noalign{\vskip1pt}
$\calB (\times 10^{-6})$ & $25.5\pm2.1^{+3.6}_{-3.9}$ &
$0.92\pm0.32\pm0.14$ &  $23.7\pm1.4\pm1.4$ \\
\noalign{\vskip1pt}
\fl\ & $0.992\pm0.024^{+0.026}_{-0.013}$ & $0.75\pm0.14\pm0.04$ & $0.950\pm0.015\pm0.006$\\
C$_L$ & $0.01\pm0.15\pm0.06$ &  $0.2\pm0.8\pm0.3$ & - \\
S$_L$ & $-0.17\pm0.20^{+0.05}_{-0.06}$ & $0.3\pm0.7\pm0.2$ & - \\
A$_{CP}$ & - & - & $-0.054\pm0.055\pm0.01$ \\
\hline
\end{tabular}
\end{center}
\end{table}

Belle has seen direct \CP\ in $\Bz\to\pip\pim$ but \babar\ does not,
reporting only that $C_{\pip\pim} = -0.25\pm0.08\pm0.02$ with a
significance of just 2.2$\sigma$. However, both experiments see
significant direct \CP\ in $\Bz\to\Kp\pim$ with \babar\ reporting
$\ACP = -0.107\pm0.016^{+0.006}_{-0.004}$ with 6.1$\sigma$
significance, to be compared to $-0.094\pm0.018\pm0.008$ from
Belle. Both experiments also measure \ACP\ for $\Bpm\to\Kpm\piz$ to be
slightly positive but consistent with zero. \ACP\ should be similar
for both $K\pi$ modes but Belle reports a 4.4$\sigma$ difference and
\babar\ sees a similar discrepancy~\cite{bib:p11}.

\end{document}